\documentclass[slac_one]{revtex4}

\usepackage{graphicx}

\usepackage{fancyhdr}
\usepackage{feynarts}
\begin{document}

\title{{\small{2005 International Linear Collider Workshop - Stanford,
U.S.A.}}\\
\vspace{12pt}
\boldmath
Dominant Two-Loop Electroweak Correction to $H\to\gamma\gamma$
\unboldmath}
\author{Bernd A. Kniehl}
\affiliation{{II.} Institut f\"ur Theoretische Physik, Universit\"at Hamburg,
Luruper Chaussee 149, 22761 Hamburg, Germany}
\begin{abstract}
We discuss a recent analysis of a dominant two-loop electroweak correction,
of $\mathcal{O}(G_{F} M_{t}^{2})$, to the partial width of the decay of an
intermediate-mass Higgs boson into a pair of photons.
The asymptotic-expansion technique was used in order to extract the leading
dependence on the top-quark mass plus four expansion terms that describe the
dependence on the $W$- and Higgs-boson masses.
This correction reduces the Born result by approximately 2.5\%.
As a by-product of this analysis, also the $\mathcal{O}(G_{F} M_{t}^{2})$
correction to the partial width of the Higgs-boson decay to two gluon jets was
recovered. 
\end{abstract}

\maketitle

\thispagestyle{fancy}

\section{\label{sec:intro}Introduction}

The Higgs boson is the missing link of the Standard Model (SM) of elementary
particle physics.
If this particle will be discovered at the Fermilab Tevatron or the CERN LHC,
then an important experimental task at a future $e^+e^-$ linear collider (ILC)
will be to determine its properties with high precision.
The electroweak precision data mainly collected at CERN LEP and SLAC SLC in
combination with the direct top-quark mass measurement at the Tevatron favour
a Higgs boson with mass $M_H=129^{+74}_{-49}$~GeV with an upper bound of about
285~GeV at the 95\% confidence level~\cite{ewwg}.
Incidentally, this $M_H$ window includes the one that is encompassed by the
vacuum-stability lower bound and the triviality upper bound allowing the SM to
be valid up to the grand-unified-theory scale $\Lambda\approx10^{16}$~GeV
\cite{ham}.
This roughly corresponds to the so-called intermediate mass range, defined by
$M_W\le M_H\le 2M_W$.
In this mass range, the decay into two photons has a branching fraction of up
to 0.3\% \cite{Kniehl:1993ay}, represents one of the most useful detection
modes at hadron colliders, and produces a clear signal at the ILC.
The cross section of $\gamma\gamma\to H$, to be measured in the $\gamma\gamma$
operation mode of the ILC, is proportional to the partial decay width
$\Gamma(H\to\gamma\gamma)$.
In conclusion, the precise knowledge of $\Gamma(H\to\gamma\gamma)$ is required
for $M_W\le M_H\le 2M_W$. 

Since there is no direct coupling of the Higgs boson to photons, the process
$H\to\gamma\gamma$ is loop-induced and so provides a handle on new charged
massive particles that are to heavy to be produced on-shell with available
particle accelerators.
The lowest-order result for $\Gamma(H\to\gamma\gamma)$ has been known for
three decades \cite{Ellis:1975ap}.
QCD corrections, which only affect the diagrams involving virtual quarks are
known at two \cite{Djouadi:1990aj} and three \cite{Ste96} loops.
In Ref.~\cite{Korner:1995xd}, the dominant two-loop electroweak correction for
a high-mass Higgs boson was found by means of the Goldstone boson equivalence
theorem.
In Ref.~\cite{Djouadi:1997rj}, the dominant two-loop electroweak correction
induced by a sequential isodoublet of ultraheavy quarks was investigated by
means of a low-energy theorem \cite{let}.
Recently, also the two-loop electroweak correction induced by light-fermion
loops has been evaluated \cite{Aglietti:2004nj}.
Here, we discuss the two-loop electroweak correction that is enhanced by
$G_F M_t^2$ \cite{fugel}.

Due to electromagnetic gauge invariance, the amputated transition-matrix
element of $H \to \gamma \gamma$ possesses the structure
\begin{eqnarray}
  \mathcal{T}^{\mu \nu} &=&
  (q_{1}\!\cdot\!q_{2} \, g^{\mu \nu} - q_{1}^{\nu}q_{2}^{\mu}) \mathcal{A},
  \label{Amplitude}
\end{eqnarray}
where $\mu$ and $\nu$ are the Lorentz indices of the external photons with
four-momenta $q_{1}$ and $q_{2}$, respectively.
Thus, we have
\begin{equation}
  \Gamma(H \to \gamma \gamma) = \frac{M_{H}^{3}}{64 \pi}|\mathcal{A}|^2.
\end{equation}
The form factor $\mathcal{A}$ is evaluated in perturbation theory as
\begin{equation}
\mathcal{A}=\mathcal{A}_{t}^{(0)} 
  + \mathcal{A}_{W}^{(0)} 
  + \mathcal{A}_{tW}^{(1)} + \cdots,
  \label{Notation}
\end{equation}
where ${\cal A}_t^{(0)}$ and ${\cal A}_W^{(0)}$ denote the one-loop 
contributions induced by virtual top quarks and $W$ bosons, respectively,
${\cal A}_{tW}^{(1)}$ stands for the two-loop electroweak correction involving
virtual top quarks, and the ellipsis represents the residual one- and two-loop
contributions as well as all contributions involving more than two loops.

Prior to discussing the results for $\mathcal{A}_{t}^{(0)}$,
$\mathcal{A}_{W}^{(0)}$, and $\mathcal{A}_{tW}^{(1)}$, we summarize the
approximations, techniques, and checks applied in Ref.~\cite{fugel}.
For simplicity, the bottom-quark mass is neglected, and the element $V_{tb}$
of the Cabibbo-Kobayashi-Maskawa quark mixing matrix is set to unity, so that
the quarks of the third fermion generation decouple from those of the first
two, which they actually do to very good approximation \cite{Hagiwara:fs}.
Exploiting the (formal) hierarchy
$M_H^2=2q_1\!\cdot\!q_2\ll(2M_W)^2\ll M_t^2$, the method of asymptotic
expansion \cite{Smirnov:pj} is applied to evaluate the results as Taylor
expansions in $\tau_t=M_H^2/(2M_t)^2$ and $\tau_W=M_H^2/(2M_W)^2$.
The on-mass-shell scheme is adopted, and the ultraviolet (UV) divergences are
regularized by means of dimensional regularization.
The anti-commuting definition of $\gamma_5$ is employed.
The tadpole contributions are treated properly.
The coefficients of the tensors $q_{1}\!\cdot\!q_{2}\,g^{\mu \nu}$ and
$q_{1}^{\nu}q_{2}^{\mu}$ in Eq.~(\ref{Amplitude}) are projected out, evaluated
separately, and found to agree.
The UV divergences are found to cancel in the final result.
The general $R_{\xi}$ gauge in adopted for the $W$ boson, and the gauge
parameter $\xi_W$ is found to drop out in the final result.
Terms quartic in $M_t$, which arise from the asymptotic expansion, the genuine
two-loop tadpoles, and the counterterms are found to cancel.
As a by-product of the two-loop calculation, the known result
\cite{Djouadi:1994ge} for the electroweak two-loop correction of
$\mathcal{O}(G_FM_t^2)$ to $\Gamma(H\to gg)$ is recovered.
Finally, the convergence properties of the expansions in $\tau_t$ and $\tau_W$
are checked.

This presentation is organized as follows.
In Sections~\ref{sec:1loop}, we illustrate the usefulness of the
asymptotic-expansion technique by redoing the one-loop calculation.
In Section~\ref{sec:2loop}, we discuss the two-loop calculation.
Section~\ref{sec:numerics} contains the discussion of the numerical results.
We conclude with a summary in Section~\ref{sec:summary}.

\section{\label{sec:1loop}One-loop results}

\begin{figure*}[ht]
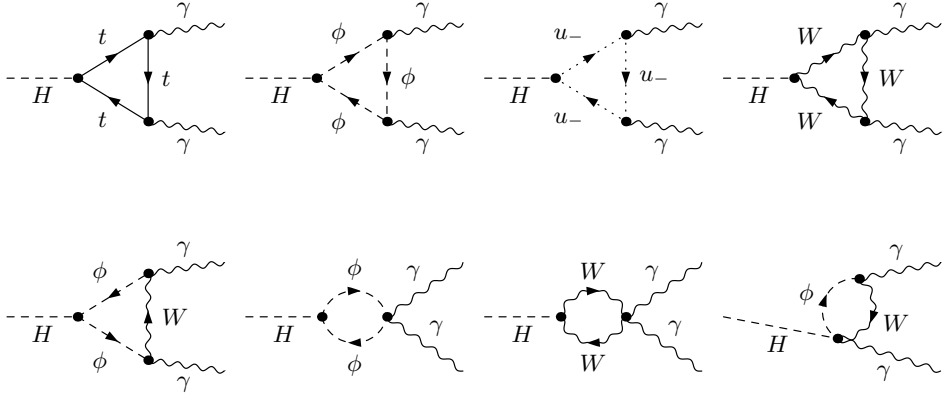

\centering
\unitlength=1bp%
\begin{feynartspicture}(370,180)(4,2)

\FADiagram{}
\FAProp(0.,10.)(6.5,10.)(0.,){/ScalarDash}{0}
\FALabel(3.25,9.18)[t]{$H$}
\FAProp(20.,15.)(13.,14.)(0.,){/Sine}{0}
\FALabel(16.2808,15.5544)[b]{$\gamma$}
\FAProp(20.,5.)(13.,6.)(0.,){/Sine}{0}
\FALabel(16.2808,4.44558)[t]{$\gamma$}
\FAProp(6.5,10.)(13.,14.)(0.,){/Straight}{1}
\FALabel(9.20801,13.1807)[br]{$t$}
\FAProp(6.5,10.)(13.,6.)(0.,){/Straight}{-1}
\FALabel(9.20801,6.81927)[tr]{$t$}
\FAProp(13.,14.)(13.,6.)(0.,){/Straight}{1}
\FALabel(14.274,10.)[l]{$t$}
\FAVert(6.5,10.){0}
\FAVert(13.,14.){0}
\FAVert(13.,6.){0}

\FADiagram{}
\FAProp(0.,10.)(6.5,10.)(0.,){/ScalarDash}{0}
\FALabel(3.25,9.18)[t]{$H$}
\FAProp(20.,15.)(13.,14.)(0.,){/Sine}{0}
\FALabel(16.2808,15.5544)[b]{$\gamma$}
\FAProp(20.,5.)(13.,6.)(0.,){/Sine}{0}
\FALabel(16.2808,4.44558)[t]{$\gamma$}
\FAProp(6.5,10.)(13.,14.)(0.,){/ScalarDash}{1}
\FALabel(9.20801,13.1807)[br]{$\phi$}
\FAProp(6.5,10.)(13.,6.)(0.,){/ScalarDash}{-1}
\FALabel(9.20801,6.81927)[tr]{$\phi$}
\FAProp(13.,14.)(13.,6.)(0.,){/ScalarDash}{1}
\FALabel(14.274,10.)[l]{$\phi$}
\FAVert(6.5,10.){0}
\FAVert(13.,14.){0}
\FAVert(13.,6.){0}

\FADiagram{}
\FAProp(0.,10.)(6.5,10.)(0.,){/ScalarDash}{0}
\FALabel(3.25,9.18)[t]{$H$}
\FAProp(20.,15.)(13.,14.)(0.,){/Sine}{0}
\FALabel(16.2808,15.5544)[b]{$\gamma$}
\FAProp(20.,5.)(13.,6.)(0.,){/Sine}{0}
\FALabel(16.2808,4.44558)[t]{$\gamma$}
\FAProp(6.5,10.)(13.,14.)(0.,){/GhostDash}{1}
\FALabel(9.20801,13.1807)[br]{$u_-$}
\FAProp(6.5,10.)(13.,6.)(0.,){/GhostDash}{-1}
\FALabel(9.20801,6.81927)[tr]{$u_-$}
\FAProp(13.,14.)(13.,6.)(0.,){/GhostDash}{1}
\FALabel(14.274,10.)[l]{$u_-$}
\FAVert(6.5,10.){0}
\FAVert(13.,14.){0}
\FAVert(13.,6.){0}

\FADiagram{}
\FAProp(0.,10.)(6.5,10.)(0.,){/ScalarDash}{0}
\FALabel(3.25,9.18)[t]{$H$}
\FAProp(20.,15.)(13.,14.)(0.,){/Sine}{0}
\FALabel(16.2808,15.5544)[b]{$\gamma$}
\FAProp(20.,5.)(13.,6.)(0.,){/Sine}{0}
\FALabel(16.2808,4.44558)[t]{$\gamma$}
\FAProp(6.5,10.)(13.,14.)(0.,){/Sine}{1}
\FALabel(9.20801,13.1807)[br]{$W$}
\FAProp(6.5,10.)(13.,6.)(0.,){/Sine}{-1}
\FALabel(9.20801,6.81927)[tr]{$W$}
\FAProp(13.,14.)(13.,6.)(0.,){/Sine}{1}
\FALabel(14.274,10.)[l]{$W$}
\FAVert(6.5,10.){0}
\FAVert(13.,14.){0}
\FAVert(13.,6.){0}

\FADiagram{}
\FAProp(0.,10.)(6.5,10.)(0.,){/ScalarDash}{0}
\FALabel(3.25,9.18)[t]{$H$}
\FAProp(20.,15.)(13.,14.)(0.,){/Sine}{0}
\FALabel(16.2808,15.5544)[b]{$\gamma$}
\FAProp(20.,5.)(13.,6.)(0.,){/Sine}{0}
\FALabel(16.2808,4.44558)[t]{$\gamma$}
\FAProp(6.5,10.)(13.,14.)(0.,){/ScalarDash}{-1}
\FALabel(9.20801,13.1807)[br]{$\phi$}
\FAProp(6.5,10.)(13.,6.)(0.,){/ScalarDash}{1}
\FALabel(9.20801,6.81927)[tr]{$\phi$}
\FAProp(13.,14.)(13.,6.)(0.,){/Sine}{-1}
\FALabel(14.274,10.)[l]{$W$}
\FAVert(6.5,10.){0}
\FAVert(13.,14.){0}
\FAVert(13.,6.){0}

\FADiagram{}
\FAProp(0.,10.)(7.,10.)(0.,){/ScalarDash}{0}
\FALabel(3.5,9.18)[t]{$H$}
\FAProp(20.,15.)(13.,10.)(0.,){/Sine}{0}
\FALabel(16.0791,13.2813)[br]{$\gamma$}
\FAProp(20.,5.)(13.,10.)(0.,){/Sine}{0}
\FALabel(16.9209,8.28129)[bl]{$\gamma$}
\FAProp(7.,10.)(13.,10.)(0.8,){/ScalarDash}{-1}
\FALabel(10.,6.53)[t]{$\phi$}
\FAProp(7.,10.)(13.,10.)(-0.8,){/ScalarDash}{1}
\FALabel(10.,13.47)[b]{$\phi$}
\FAVert(7.,10.){0}
\FAVert(13.,10.){0}

\FADiagram{}
\FAProp(0.,10.)(7.,10.)(0.,){/ScalarDash}{0}
\FALabel(3.5,9.18)[t]{$H$}
\FAProp(20.,15.)(13.,10.)(0.,){/Sine}{0}
\FALabel(16.0791,13.2813)[br]{$\gamma$}
\FAProp(20.,5.)(13.,10.)(0.,){/Sine}{0}
\FALabel(16.9209,8.28129)[bl]{$\gamma$}
\FAProp(7.,10.)(13.,10.)(0.8,){/Sine}{-1}
\FALabel(10.,6.53)[t]{$W$}
\FAProp(7.,10.)(13.,10.)(-0.8,){/Sine}{1}
\FALabel(10.,13.47)[b]{$W$}
\FAVert(7.,10.){0}
\FAVert(13.,10.){0}

\FADiagram{}
\FAProp(0.,10.)(10.5,8.)(0.,){/ScalarDash}{0}
\FALabel(5.00675,8.20296)[t]{$H$}
\FAProp(20.,15.)(12.5,13.5)(0.,){/Sine}{0}
\FALabel(15.946,15.2899)[b]{$\gamma$}
\FAProp(20.,5.)(10.5,8.)(0.,){/Sine}{0}
\FALabel(14.7832,5.50195)[t]{$\gamma$}
\FAProp(12.5,13.5)(10.5,8.)(0.8,){/ScalarDash}{-1}
\FALabel(8.32332,12.0797)[r]{$\phi$}
\FAProp(12.5,13.5)(10.5,8.)(-0.8,){/Sine}{1}
\FALabel(14.6767,9.4203)[l]{$W$}
\FAVert(12.5,13.5){0}
\FAVert(10.5,8.){0}

\end{feynartspicture}
\caption{Typical one-loop diagrams contributing to $H\to\gamma\gamma$.}
\label{1loop}
\end{figure*}

Typical Feynman diagrams contributing at one loop in $R_\xi$ gauge are
depicted in Fig.~\ref{1loop}, where $\phi$ and $u$ denote the charged
Goldstone bosons and Faddeev-Popov ghosts, respectively.
The analytic expression for $\mathcal{A}_{t}^{(0)}$ and
$\mathcal{A}_{W}^{(0)}$ in Eq.~(\ref{Notation}) and their expansions in
$\tau_t$ and $\tau_W$ read:
\begin{eqnarray}
  \mathcal{A}_{t}^{(0)} & = & 
  \hat{\mathcal{A}} N_c Q_t^2 \left\{ \frac{1}{\tau_{t}} \left[ 1 +
  \left( 1 - \frac{1}{\tau_{t}} \right) \arcsin^{2}\sqrt{\tau_{t}}
  \right] \right\}
  \nonumber\\ 
  & = & \hat{\mathcal{A}} N_c Q_t^2 \left( \frac{2}{3} + \frac{7}{45} \tau_{t}
  + \frac{4}{63} \tau_{t}^{2} 
  + \frac{52}{1575} \tau_{t}^{3} 
  + \frac{1024}{51975} \tau_{t}^{4} 
  + \frac{2432}{189189} \tau_{t}^{5} 
  + \ldots \right),
  \\ 
  \mathcal{A}_{W}^{(0)} & = & 
  \hat{\mathcal{A}} \left\{ - \frac{1}{2} \left[ 2 +
  \frac{3}{\tau_{W}} + \frac{3}{\tau_{W}} \left( 2 -
  \frac{1}{\tau_{W}} \right) \arcsin^{2}\sqrt{\tau_{W}} \right]
  \right\}
  \nonumber\\
  & = & \hat{\mathcal{A}} \left( - \frac{7}{2} -
  \frac{11}{15} \tau_{W} 
  - \frac{38}{105} \tau_{W}^{2} 
  - \frac{116}{525} \tau_{W}^{3} 
  - \frac{2624}{17325} \tau_{W}^{4} 
  - \frac{640}{5733} \tau_{W}^{5} 
  + \ldots \right),
\label{eq:born}
\end{eqnarray}
where $\hat{\mathcal{A}} = 2^{1/4} G_{F}^{1/2}(\alpha/\pi)$.
Here, $\alpha$ is Sommerfeld's fine-structure constant, $G_F$ is Fermi's
constant, $N_{c}=3$ is the number of quark colours, and $Q_{t}=2/3$ is the
electric charge of the top quark in units of the positron charge.

\section{\label{sec:2loop}Two-loop results}

\begin{figure*}[ht]
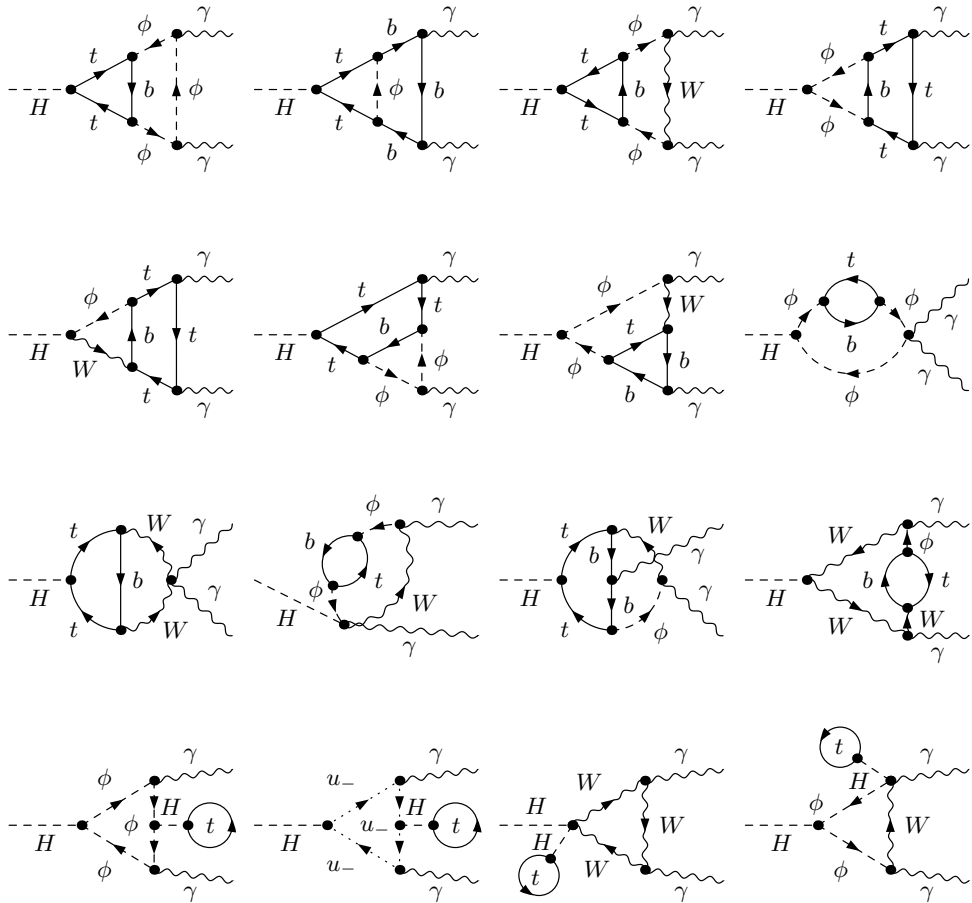

\centering
\unitlength=1bp%
\begin{feynartspicture}(370,400)(4,4)

\FADiagram{}
\FAProp(0.,10.)(5.5,10.)(0.,){/ScalarDash}{0}
\FALabel(2.75,9.18)[t]{$H$}
\FAProp(20.,15.)(15.,15.)(0.,){/Sine}{0}
\FALabel(17.5,16.07)[b]{$\gamma$}
\FAProp(20.,5.)(15.,5.)(0.,){/Sine}{0}
\FALabel(17.5,3.93)[t]{$\gamma$}
\FAProp(11.,7.1)(5.5,10.)(0.,){/Straight}{1}
\FALabel(8.00707,7.65892)[tr]{$t$}
\FAProp(11.,7.1)(15.,5.)(0.,){/ScalarDash}{1}
\FALabel(12.7595,5.15763)[tr]{$\phi$}
\FAProp(11.,12.95)(11.,7.1)(0.,){/Straight}{1}
\FALabel(12.07,10.025)[l]{$b$}
\FAProp(11.,12.95)(5.5,10.)(0.,){/Straight}{-1}
\FALabel(7.99737,12.3609)[br]{$t$}
\FAProp(11.,12.95)(15.,15.)(0.,){/ScalarDash}{-1}
\FALabel(12.7731,14.8744)[br]{$\phi$}
\FAProp(15.,15.)(15.,5.)(0.,){/ScalarDash}{-1}
\FALabel(16.07,10.)[l]{$\phi$}
\FAVert(11.,12.95){0}
\FAVert(11.,7.1){0}
\FAVert(5.5,10.){0}
\FAVert(15.,15.){0}
\FAVert(15.,5.){0}

\FADiagram{}
\FAProp(0.,10.)(5.5,10.)(0.,){/ScalarDash}{0}
\FALabel(2.75,9.18)[t]{$H$}
\FAProp(20.,15.)(15.,15.)(0.,){/Sine}{0}
\FALabel(17.5,16.07)[b]{$\gamma$}
\FAProp(20.,5.)(15.,5.)(0.,){/Sine}{0}
\FALabel(17.5,3.93)[t]{$\gamma$}
\FAProp(11.,7.1)(5.5,10.)(0.,){/Straight}{1}
\FALabel(8.00707,7.65892)[tr]{$t$}
\FAProp(11.,7.1)(15.,5.)(0.,){/Straight}{-1}
\FALabel(12.7595,5.15763)[tr]{$b$}
\FAProp(11.,12.95)(11.,7.1)(0.,){/ScalarDash}{-1}
\FALabel(12.07,10.025)[l]{$\phi$}
\FAProp(11.,12.95)(5.5,10.)(0.,){/Straight}{-1}
\FALabel(7.99737,12.3609)[br]{$t$}
\FAProp(11.,12.95)(15.,15.)(0.,){/Straight}{1}
\FALabel(12.7731,14.8744)[br]{$b$}
\FAProp(15.,15.)(15.,5.)(0.,){/Straight}{1}
\FALabel(16.07,10.)[l]{$b$}
\FAVert(11.,12.95){0}
\FAVert(11.,7.1){0}
\FAVert(5.5,10.){0}
\FAVert(15.,15.){0}
\FAVert(15.,5.){0}

\FADiagram{}
\FAProp(0.,10.)(5.5,10.)(0.,){/ScalarDash}{0}
\FALabel(2.75,9.18)[t]{$H$}
\FAProp(20.,15.)(15.,15.)(0.,){/Sine}{0}
\FALabel(17.5,16.07)[b]{$\gamma$}
\FAProp(20.,5.)(15.,5.)(0.,){/Sine}{0}
\FALabel(17.5,3.93)[t]{$\gamma$}
\FAProp(11.,7.1)(5.5,10.)(0.,){/Straight}{-1}
\FALabel(8.00707,7.65892)[tr]{$t$}
\FAProp(11.,7.1)(15.,5.)(0.,){/ScalarDash}{-1}
\FALabel(12.7595,5.15763)[tr]{$\phi$}
\FAProp(11.,12.95)(11.,7.1)(0.,){/Straight}{-1}
\FALabel(12.07,10.025)[l]{$b$}
\FAProp(11.,12.95)(5.5,10.)(0.,){/Straight}{1}
\FALabel(7.99737,12.3609)[br]{$t$}
\FAProp(11.,12.95)(15.,15.)(0.,){/ScalarDash}{1}
\FALabel(12.7731,14.8744)[br]{$\phi$}
\FAProp(15.,15.)(15.,5.)(0.,){/Sine}{1}
\FALabel(16.07,10.)[l]{$W$}
\FAVert(11.,12.95){0}
\FAVert(11.,7.1){0}
\FAVert(5.5,10.){0}
\FAVert(15.,15.){0}
\FAVert(15.,5.){0}

\FADiagram{}
\FAProp(0.,10.)(5.5,10.)(0.,){/ScalarDash}{0}
\FALabel(2.75,9.18)[t]{$H$}
\FAProp(20.,15.)(15.,15.)(0.,){/Sine}{0}
\FALabel(17.5,16.07)[b]{$\gamma$}
\FAProp(20.,5.)(15.,5.)(0.,){/Sine}{0}
\FALabel(17.5,3.93)[t]{$\gamma$}
\FAProp(11.,7.1)(5.5,10.)(0.,){/ScalarDash}{-1}
\FALabel(8.00707,7.65892)[tr]{$\phi$}
\FAProp(11.,7.1)(15.,5.)(0.,){/Straight}{-1}
\FALabel(12.7595,5.15763)[tr]{$t$}
\FAProp(11.,12.95)(11.,7.1)(0.,){/Straight}{-1}
\FALabel(12.07,10.025)[l]{$b$}
\FAProp(11.,12.95)(5.5,10.)(0.,){/ScalarDash}{1}
\FALabel(7.99737,12.3609)[br]{$\phi$}
\FAProp(11.,12.95)(15.,15.)(0.,){/Straight}{1}
\FALabel(12.7731,14.8744)[br]{$t$}
\FAProp(15.,15.)(15.,5.)(0.,){/Straight}{1}
\FALabel(16.07,10.)[l]{$t$}
\FAVert(11.,12.95){0}
\FAVert(11.,7.1){0}
\FAVert(5.5,10.){0}
\FAVert(15.,15.){0}
\FAVert(15.,5.){0}

\FADiagram{}
\FAProp(0.,10.)(5.5,10.)(0.,){/ScalarDash}{0}
\FALabel(2.75,9.18)[t]{$H$}
\FAProp(20.,15.)(15.,15.)(0.,){/Sine}{0}
\FALabel(17.5,16.07)[b]{$\gamma$}
\FAProp(20.,5.)(15.,5.)(0.,){/Sine}{0}
\FALabel(17.5,3.93)[t]{$\gamma$}
\FAProp(11.,7.1)(5.5,10.)(0.,){/Sine}{-1}
\FALabel(8.00707,7.65892)[tr]{$W$}
\FAProp(11.,7.1)(15.,5.)(0.,){/Straight}{-1}
\FALabel(12.7595,5.15763)[tr]{$t$}
\FAProp(11.,12.95)(11.,7.1)(0.,){/Straight}{-1}
\FALabel(12.07,10.025)[l]{$b$}
\FAProp(11.,12.95)(5.5,10.)(0.,){/ScalarDash}{1}
\FALabel(7.99737,12.3609)[br]{$\phi$}
\FAProp(11.,12.95)(15.,15.)(0.,){/Straight}{1}
\FALabel(12.7731,14.8744)[br]{$t$}
\FAProp(15.,15.)(15.,5.)(0.,){/Straight}{1}
\FALabel(16.07,10.)[l]{$t$}
\FAVert(11.,12.95){0}
\FAVert(11.,7.1){0}
\FAVert(5.5,10.){0}
\FAVert(15.,15.){0}
\FAVert(15.,5.){0}

\FADiagram{}
\FAProp(0.,10.)(5.5,10.)(0.,){/ScalarDash}{0}
\FALabel(2.75,9.18)[t]{$H$}
\FAProp(20.,15.)(15.,15.)(0.,){/Sine}{0}
\FALabel(17.5,16.07)[b]{$\gamma$}
\FAProp(20.,5.)(15.,5.)(0.,){/Sine}{0}
\FALabel(17.5,3.93)[t]{$\gamma$}
\FAProp(5.5,10.)(15.,15.)(0.,){/Straight}{1}
\FALabel(10.0081,13.3916)[br]{$t$}
\FAProp(9.7,7.75)(5.5,10.)(0.,){/Straight}{1}
\FALabel(7.34806,7.98871)[tr]{$t$}
\FAProp(9.7,7.75)(15.,5.)(0.,){/ScalarDash}{1}
\FALabel(12.1161,5.47918)[tr]{$\phi$}
\FAProp(15.,10.5)(9.7,7.75)(0.,){/Straight}{1}
\FALabel(12.1161,10.0208)[br]{$b$}
\FAProp(15.,10.5)(15.,15.)(0.,){/Straight}{-1}
\FALabel(16.07,12.75)[l]{$t$}
\FAProp(15.,10.5)(15.,5.)(0.,){/ScalarDash}{-1}
\FALabel(16.07,7.75)[l]{$\phi$}
\FAVert(15.,10.5){0}
\FAVert(9.7,7.75){0}
\FAVert(5.5,10.){0}
\FAVert(15.,15.){0}
\FAVert(15.,5.){0}

\FADiagram{}
\FAProp(0.,10.)(5.5,10.)(0.,){/ScalarDash}{0}
\FALabel(2.75,9.18)[t]{$H$}
\FAProp(20.,15.)(15.,15.)(0.,){/Sine}{0}
\FALabel(17.5,16.07)[b]{$\gamma$}
\FAProp(20.,5.)(15.,5.)(0.,){/Sine}{0}
\FALabel(17.5,3.93)[t]{$\gamma$}
\FAProp(5.5,10.)(15.,15.)(0.,){/ScalarDash}{1}
\FALabel(10.0081,13.3916)[br]{$\phi$}
\FAProp(9.7,7.75)(5.5,10.)(0.,){/ScalarDash}{1}
\FALabel(7.34806,7.98871)[tr]{$\phi$}
\FAProp(9.7,7.75)(15.,5.)(0.,){/Straight}{-1}
\FALabel(12.1161,5.47918)[tr]{$b$}
\FAProp(15.,10.5)(9.7,7.75)(0.,){/Straight}{-1}
\FALabel(12.1161,10.0208)[br]{$t$}
\FAProp(15.,10.5)(15.,15.)(0.,){/Sine}{-1}
\FALabel(16.07,12.75)[l]{$W$}
\FAProp(15.,10.5)(15.,5.)(0.,){/Straight}{1}
\FALabel(16.07,7.75)[l]{$b$}
\FAVert(15.,10.5){0}
\FAVert(9.7,7.75){0}
\FAVert(5.5,10.){0}
\FAVert(15.,15.){0}
\FAVert(15.,5.){0}

\FADiagram{}
\FAProp(0.,10.)(4.5,10.)(0.,){/ScalarDash}{0}
\FALabel(2.25,9.18)[t]{$H$}
\FAProp(20.,15.)(14.6,10.)(0.,){/Sine}{0}
\FALabel(17.8731,11.8427)[tl]{$\gamma$}
\FAProp(20.,5.)(14.6,10.)(0.,){/Sine}{0}
\FALabel(16.7269,6.84267)[tr]{$\gamma$}
\FAProp(7.,13.)(4.5,10.)(0.211476,){/ScalarDash}{-1}
\FALabel(4.72204,12.2766)[br]{$\phi$}
\FAProp(12.,13.)(7.,13.)(0.8,){/Straight}{1}
\FALabel(9.5,16.07)[b]{$t$}
\FAProp(12.,13.)(7.,13.)(-0.8,){/Straight}{-1}
\FALabel(9.5,9.93)[t]{$b$}
\FAProp(12.,13.)(14.6,10.)(-0.182242,){/ScalarDash}{1}
\FALabel(14.2647,12.2721)[bl]{$\phi$}
\FAProp(4.5,10.)(14.6,10.)(0.693069,){/ScalarDash}{-1}
\FALabel(9.55,5.43)[t]{$\phi$}
\FAVert(12.,13.){0}
\FAVert(7.,13.){0}
\FAVert(4.5,10.){0}
\FAVert(14.6,10.){0}

\FADiagram{}
\FAProp(0.,10.)(5.5,10.)(0.,){/ScalarDash}{0}
\FALabel(2.75,9.18)[t]{$H$}
\FAProp(20.,15.)(14.5,10.)(0.,){/Sine}{0}
\FALabel(17.6874,14.1669)[br]{$\gamma$}
\FAProp(20.,5.)(14.5,10.)(0.,){/Sine}{0}
\FALabel(17.8126,8.16691)[bl]{$\gamma$}
\FAProp(10.,14.5)(5.5,10.)(0.4,){/Straight}{-1}
\FALabel(6.23398,13.766)[br]{$t$}
\FAProp(10.,5.5)(10.,14.5)(0.,){/Straight}{-1}
\FALabel(11.07,10.)[l]{$b$}
\FAProp(10.,5.5)(5.5,10.)(-0.4,){/Straight}{1}
\FALabel(6.23398,6.23398)[tr]{$t$}
\FAProp(10.,5.5)(14.5,10.)(0.4,){/Sine}{1}
\FALabel(13.766,6.23398)[tl]{$W$}
\FAProp(10.,14.5)(14.5,10.)(-0.4,){/Sine}{-1}
\FALabel(12.266,14.366)[bl]{$W$}
\FAVert(10.,5.5){0}
\FAVert(10.,14.5){0}
\FAVert(5.5,10.){0}
\FAVert(14.5,10.){0}

\FADiagram{}
\FAProp(0.,10.)(8.,6.)(0.,){/ScalarDash}{0}
\FALabel(3.89862,7.31724)[tr]{$H$}
\FAProp(20.,15.)(13.,15.)(0.,){/Sine}{0}
\FALabel(16.5,16.07)[b]{$\gamma$}
\FAProp(20.,5.)(8.,6.)(0.,){/Sine}{0}
\FALabel(13.8713,4.43535)[t]{$\gamma$}
\FAProp(9.2,13.9)(13.,15.)(-0.150337,){/ScalarDash}{-1}
\FALabel(10.5863,15.7445)[b]{$\phi$}
\FAProp(7.,9.5)(9.2,13.9)(0.8,){/Straight}{1}
\FALabel(10.7664,10.6068)[tl]{$t$}
\FAProp(7.,9.5)(9.2,13.9)(-0.8,){/Straight}{-1}
\FALabel(5.43364,12.7932)[br]{$b$}
\FAProp(7.,9.5)(8.,6.)(0.222749,){/ScalarDash}{1}
\FALabel(6.13962,9.00132)[r]{$\phi$}
\FAProp(13.,15.)(8.,6.)(-0.605305,){/Sine}{-1}
\FALabel(14.0988,8.71399)[tl]{$W$}
\FAVert(7.,9.5){0}
\FAVert(9.2,13.9){0}
\FAVert(13.,15.){0}
\FAVert(8.,6.){0}

\FADiagram{}
\FAProp(0.,10.)(5.5,10.)(0.,){/ScalarDash}{0}
\FALabel(2.75,9.18)[t]{$H$}
\FAProp(20.,15.)(10.,10.)(0.,){/Sine}{0}
\FALabel(17.1318,12.7564)[tl]{$\gamma$}
\FAProp(20.,5.)(14.5,10.)(0.,){/Sine}{0}
\FALabel(17.8126,8.16691)[bl]{$\gamma$}
\FAProp(10.,5.5)(5.5,10.)(-0.4,){/Straight}{1}
\FALabel(6.23398,6.23398)[tr]{$t$}
\FAProp(10.,5.5)(10.,10.)(0.,){/Straight}{-1}
\FALabel(11.07,7.75)[l]{$b$}
\FAProp(10.,5.5)(14.5,10.)(0.4,){/ScalarDash}{1}
\FALabel(13.766,6.23398)[tl]{$\phi$}
\FAProp(10.,14.5)(5.5,10.)(0.4,){/Straight}{-1}
\FALabel(6.23398,13.766)[br]{$t$}
\FAProp(10.,14.5)(10.,10.)(0.,){/Straight}{1}
\FALabel(8.93,12.25)[r]{$b$}
\FAProp(10.,14.5)(14.5,10.)(-0.4,){/Sine}{-1}
\FALabel(13.1968,14.1968)[bl]{$W$}
\FAVert(10.,14.5){0}
\FAVert(10.,5.5){0}
\FAVert(5.5,10.){0}
\FAVert(10.,10.){0}
\FAVert(14.5,10.){0}

\FADiagram{}
\FAProp(0.,10.)(5.5,10.)(0.,){/ScalarDash}{0}
\FALabel(2.75,9.18)[t]{$H$}
\FAProp(20.,15.)(14.5,15.)(0.,){/Sine}{0}
\FALabel(17.25,16.07)[b]{$\gamma$}
\FAProp(20.,5.)(14.5,5.)(0.,){/Sine}{0}
\FALabel(17.25,3.93)[t]{$\gamma$}
\FAProp(5.5,10.)(14.5,15.)(0.,){/Sine}{-1}
\FALabel(9.72725,13.3749)[br]{$W$}
\FAProp(5.5,10.)(14.5,5.)(0.,){/Sine}{1}
\FALabel(9.72725,6.62506)[tr]{$W$}
\FAProp(14.5,12.5)(14.5,15.)(0.,){/ScalarDash}{1}
\FALabel(15.57,13.30)[l]{$\phi$}
\FAProp(14.5,7.5)(14.5,12.5)(0.8,){/Straight}{-1}
\FALabel(17.57,10.)[l]{$t$}
\FAProp(14.5,7.5)(14.5,12.5)(-0.8,){/Straight}{1}
\FALabel(11.43,10.)[r]{$b$}
\FAProp(14.5,7.5)(14.5,5.)(0.,){/Sine}{-1}
\FALabel(15.57,6.50)[l]{$W$}
\FAVert(14.5,7.5){0}
\FAVert(14.5,12.5){0}
\FAVert(5.5,10.){0}
\FAVert(14.5,15.){0}
\FAVert(14.5,5.){0}

\FADiagram{}
\FAProp(0.,10.)(6.5,10.)(0.,){/ScalarDash}{0}
\FALabel(3.25,9.18)[t]{$H$}
\FAProp(20.,15.)(13.,14.)(0.,){/Sine}{0}
\FALabel(16.2808,15.5544)[b]{$\gamma$}
\FAProp(20.,5.)(13.,6.)(0.,){/Sine}{0}
\FALabel(16.2808,4.44558)[t]{$\gamma$}
\FAProp(6.5,10.)(13.,14.)(0.,){/ScalarDash}{1}
\FALabel(9.20801,13.1807)[br]{$\phi$}
\FAProp(6.5,10.)(13.,6.)(0.,){/ScalarDash}{-1}
\FALabel(9.20801,6.81927)[tr]{$\phi$}
\FAProp(13.,14.)(13.,10.)(0.,){/ScalarDash}{1}
\FALabel(10.274,10.)[l]{$\phi$}
\FAProp(13.,10.)(13.,6.)(0.,){/ScalarDash}{1}
\FAProp(13.,10.)(16.,10.)(0.,){/ScalarDash}{0}
\FALabel(13.500,11.7)[l]{$H$}
\FAProp(16.,10.)(16.,10.)(20.,10.){/Straight}{1}
\FALabel(17.674,10.)[l]{$t$}
\FAVert(6.5,10.){0}
\FAVert(13.,14.){0}
\FAVert(13.,6.){0}
\FAVert(16.,10.){0}
\FAVert(13.,10.){0}

\FADiagram{}
\FAProp(0.,10.)(6.5,10.)(0.,){/ScalarDash}{0}
\FALabel(3.25,9.18)[t]{$H$}
\FAProp(20.,15.)(13.,14.)(0.,){/Sine}{0}
\FALabel(16.2808,15.5544)[b]{$\gamma$}
\FAProp(20.,5.)(13.,6.)(0.,){/Sine}{0}
\FALabel(16.2808,4.44558)[t]{$\gamma$}
\FAProp(6.5,10.)(13.,14.)(0.,){/GhostDash}{1}
\FALabel(9.20801,13.1807)[br]{$u_-$}
\FAProp(6.5,10.)(13.,6.)(0.,){/GhostDash}{-1}
\FALabel(9.20801,6.81927)[tr]{$u_-$}
\FAProp(13.,14.)(13.,10.)(0.,){/GhostDash}{1}
\FALabel(9.474,9.7)[l]{$u_-$}
\FAProp(13.,10.)(13.,6.)(0.,){/GhostDash}{1}
\FAProp(13.,10.)(16.,10.)(0.,){/ScalarDash}{0}
\FALabel(13.500,11.7)[l]{$H$}
\FAProp(16.,10.)(16.,10.)(20.,10.){/Straight}{1}
\FALabel(17.674,10.)[l]{$t$}
\FAVert(6.5,10.){0}
\FAVert(13.,14.){0}
\FAVert(13.,6.){0}
\FAVert(16.,10.){0}
\FAVert(13.,10.){0}

\FADiagram{}
\FAProp(0.,10.)(6.5,10.)(0.,){/ScalarDash}{0}
\FALabel(3.25,12.5)[t]{$H$}
\FAProp(6.5,10.)(4.5,7.)(0.,){/ScalarDash}{0}
\FALabel(4.80,8.5)[r]{$H$}
\FAProp(4.5,7.)(4.5,7.)(2.5,4.){/Straight}{1}
\FALabel(2.8,5.3)[l]{$t$}
\FAProp(20.,15.)(13.,14.)(0.,){/Sine}{0}
\FALabel(16.2808,15.5544)[b]{$\gamma$}
\FAProp(20.,5.)(13.,6.)(0.,){/Sine}{0}
\FALabel(16.2808,4.44558)[t]{$\gamma$}
\FAProp(6.5,10.)(13.,14.)(0.,){/Sine}{1}
\FALabel(9.20801,13.1807)[br]{$W$}
\FAProp(6.5,10.)(13.,6.)(0.,){/Sine}{-1}
\FALabel(9.90801,6.81927)[tr]{$W$}
\FAProp(13.,14.)(13.,6.)(0.,){/Sine}{1}
\FALabel(14.274,10.)[l]{$W$}
\FAVert(6.5,10.){0}
\FAVert(13.,14.){0}
\FAVert(13.,6.){0}
\FAVert(6.5,10.){0}
\FAVert(4.5,7.){0}

\FADiagram{}
\FAProp(0.,10.)(6.5,10.)(0.,){/ScalarDash}{0}
\FALabel(3.25,9.18)[t]{$H$}
\FAProp(20.,15.)(13.,14.)(0.,){/Sine}{0}
\FALabel(16.2808,15.5544)[b]{$\gamma$}
\FAProp(13.,14.)(10.,16.)(0.,){/ScalarDash}{0}
\FALabel(11.20,14.)[r]{$H$}
\FAProp(10.,16.)(10.,16.)(7.,18.){/Straight}{1}
\FALabel(8.0,17.)[l]{$t$}
\FAProp(20.,5.)(13.,6.)(0.,){/Sine}{0}
\FALabel(16.2808,4.44558)[t]{$\gamma$}
\FAProp(6.5,10.)(13.,14.)(0.,){/ScalarDash}{-1}
\FALabel(7.00801,10.9807)[br]{$\phi$}
\FAProp(6.5,10.)(13.,6.)(0.,){/ScalarDash}{1}
\FALabel(9.20801,6.81927)[tr]{$\phi$}
\FAProp(13.,14.)(13.,6.)(0.,){/Sine}{-1}
\FALabel(14.274,10.)[l]{$W$}
\FAVert(6.5,10.){0}
\FAVert(13.,14.){0}
\FAVert(13.,6.){0}
\FAVert(13.,14.){0}
\FAVert(10.,16.){0}

\end{feynartspicture}
\caption{Typical two-loop electroweak diagrams contributing to
$H\to\gamma\gamma$.}
\label{2loop}
\end{figure*}

The contributions of $\mathcal{O}(G_F M_t^2)$ are obtained by considering all
two-loop electroweak diagrams involving a virtual top quark.
This includes also the tadpole diagrams with a closed top-quark loop, which
are proportional to $M_t^4$.
For arbitrary gauge parameter, this leads us to consider a total of order 1000
diagrams.
Some of them are depicted in Fig.~\ref{2loop}.
These diagrams naturally split into two classes.
The first class consists of those diagrams where a neutral boson, i.e.\ a
Higgs boson or a neutral Goldstone boson ($\chi$), is added to the one-loop
top-quark diagrams.
The exchange of a $Z$ boson does not produce quadratic contributions in $M_t$.
The application of the asymptotic-expansion technique to these diagrams leads
to a simple Taylor expansion in the external momenta.
This is different for the second class of diagrams, which, next to the top
quark, also contain a $W$ or a $\phi$ boson and, as a consequence, also the
bottom quark, which is taken to be massless throughout the calculation. 
Due to the presence of cuts through light-particle lines, the
asymptotic-expansion technique applied to these diagrams also yields
nontrivial terms.

The final result for ${\cal A}_{tW}^{(1)}$ emerges as the sum
\begin{equation}
  \mathcal{A}_{tW}^{(1)} = 
  \mathcal{A}_{u}^{(1)} +
  \mathcal{A}_{H,\chi}^{(1)} +
  \mathcal{A}_{W,\phi}^{(1)},
\label{eq:sum}
\end{equation}
where $\mathcal{A}_{u}^{(1)}$ is the universal contribution induced by the
renormalization of the Higgs-boson wave function and the factor $1/M_W$ common
to all one-loop diagrams \cite{hll}, $\mathcal{A}_{H,\chi}^{(1)}$ is the
two-loop contribution involving virtual $H$ and $\chi$ bosons, and
$\mathcal{A}_{W,\phi}^{(1)}$ the remaining two-loop contribution involving
virtual $W$ and $\phi$ bosons.
In $\mathcal{A}_{H,\chi}^{(1)}$ and $\mathcal{A}_{W,\phi}^{(1)}$, also
the corresponding counterterm and tadpole contributions are included.
For the individual pieces, one has \cite{fugel}
\begin{eqnarray}
  \mathcal{A}_{u}^{(1)} & = & 
  \hat{\mathcal{A}} N_c x_t \left( - \frac{329}{108} 
    - \frac{77}{90} \tau_{W} - \frac{19}{45} \tau_{W}^{2} 
    - \frac{58}{225} \tau_{W}^{3} - \frac{1312}{7425} \tau_{W}^{4} 
    +\cdots
  \right) ,
  \nonumber\\
  \mathcal{A}_{H,\chi}^{(1)} & = & 
  \hat{\mathcal{A}} N_c x_t \left( - \frac{8}{27} \right) ,
  \nonumber\\
  \mathcal{A}_{W,\phi}^{(1)} & = & 
  \hat{\mathcal{A}} N_c x_t \left( \frac{182}{27} + \frac{22}{15} \tau_{W} +
    \frac{76}{105} \tau_{W}^{2} + \frac{232}{525} \tau_{W}^{3} 
    + \frac{5248}{17325} \tau_{W}^{4} +\cdots
  \right) ,
\label{eq:terms}
\end{eqnarray}
where $\hat{\mathcal{A}}$ is defined below Eq.~(\ref{eq:born}),
$x_t=G_FM_t^2/(8\pi^2\sqrt{2})$, and the ellipses indicate terms of
$\mathcal{O}(\tau_W^5)$.
Notice that the leading $\mathcal{O}(G_F M_t^2)$ term of
$\mathcal{A}_{H,\chi}^{(1)}$ is not accompanied by an expansion in $\tau_W$,
since the contributing diagrams do not involve virtual $W$ or $\phi$ bosons.
On the other hand, detailed inspection reveals that there is also no 
expansion in the parameter $M_H^2/(2M_Z)^2$, contrary to what might be
expected at first sight.
Inserting Eq.~(\ref{eq:terms}) into Eq.~(\ref{eq:sum}), one obtains the final
result
\begin{equation}
  \mathcal{A}_{tW}^{(1)} = 
  \hat{\mathcal{A}} N_c x_t \left( \frac{367}{108} + \frac{11}{18} \tau_{W} +
  \frac{19}{63} \tau_{W}^{2} + \frac{58}{315} \tau_{W}^{3} 
  + \frac{1312}{10395} \tau_{W}^{4} +\cdots
\right).
\label{eq:end}
\end{equation}

\section{\label{sec:numerics}Numerical results}

We are now in a position to discuss the numerical results.
They are evaluated using the following numerical values for the input
parameters: 
$G_F=1.16639\times10^{-5}$~GeV$^{-2}$,
$M_W=80.423$~GeV, and
$M_t=174.3$~GeV \cite{Hagiwara:fs}.

\begin{figure*}[ht]
\centering
\begin{tabular}{cc}
\includegraphics[width=90mm]{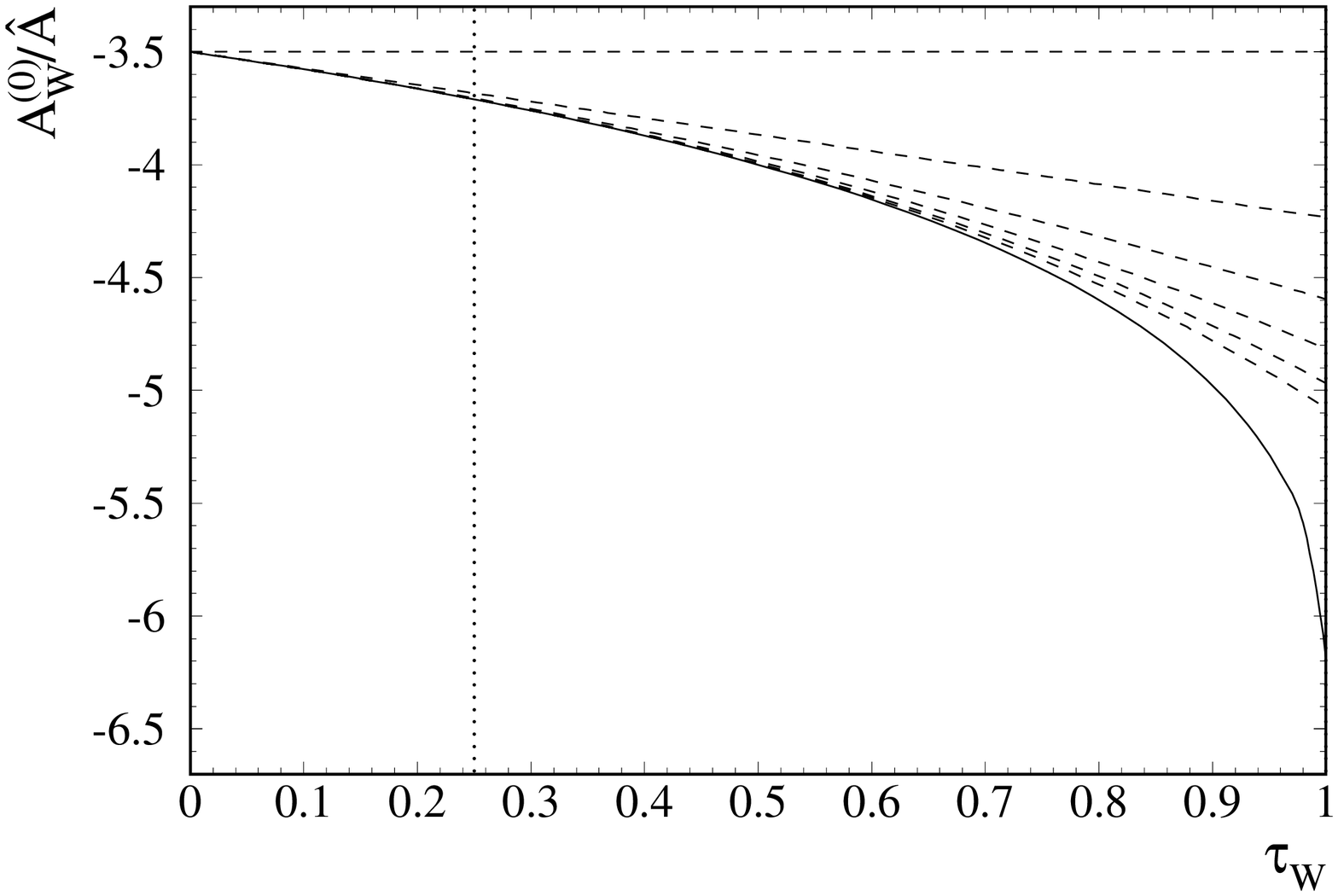} &
\includegraphics[width=90mm]{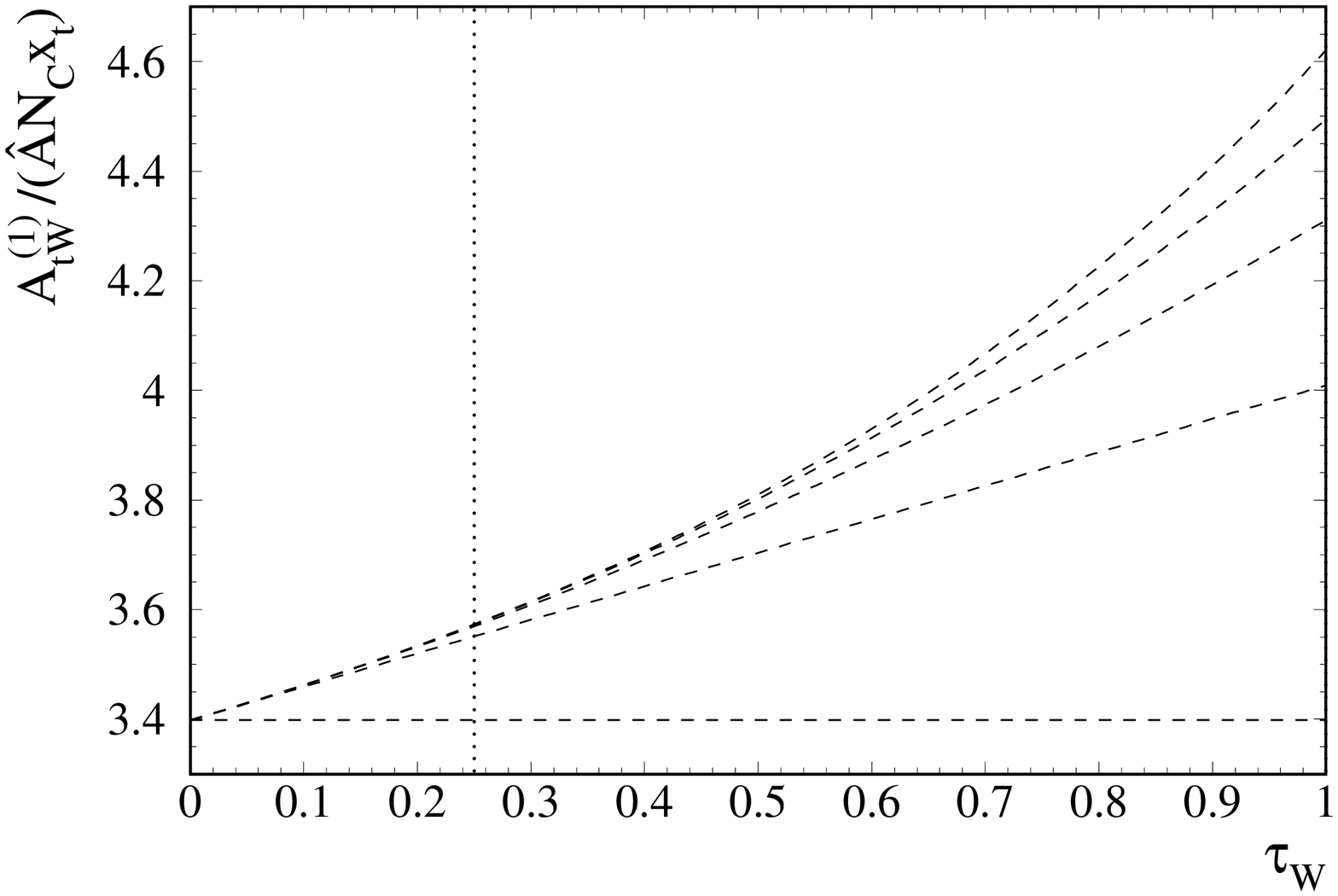} \\
(a) & (b)
\end{tabular}
\caption{(a) $\mathcal{A}_W^{(0)}$ normalized to $\hat{\mathcal{A}}$ and (b)
$\mathcal{A}_{tW}^{(1)}$ normalized to $\hat{\mathcal{A}}N_cx_t$ as functions
of $\tau_W$.}
\label{loopW}
\end{figure*}

We first assess the convergence property of the $\tau_W$ expansion of 
$\mathcal{A}_{tW}^{(1)}$ in Eq.~(\ref{eq:end}).
To this end, it is useful to also consider the example of ${\cal A}_W^{(0)}$,
for which the exact result is known.
In Figs.~\ref{loopW}(a) and (b), ${\cal A}_W^{(0)}$ and
$\mathcal{A}_{tW}^{(1)}$ are presented as functions of $\tau_{W}$,
respectively.
The dashed curves represent the sequences of approximations that are obtained
by successively including higher powers of $\tau_{W}$ in the expansions, while
the solid curve in Fig.~\ref{loopW}(a) indicates the exact result.
The dotted vertical lines and the right edges of the frames encompass the
$M_H$ range $M_W\le M_H\le 2M_W$.
We observe from Fig.~\ref{loopW}(a) that, for $M_H=120$~GeV, 140~GeV, and
$2M_W$, the approximation for ${\cal A}_W^{(0)}$ by five expansion terms
deviates from the exact result by as little as 0.3\%, 1.6\%, and 19.9\%,
respectively.
The relatively modest description towards $M_H=2M_W$, i.e.\ $\tau_W=1$, may be
understood by observing that the exact result behaves like $\sqrt{1-\tau_{W}}$
in this limit.
From Fig.~\ref{loopW}(b), we see that the $\tau_W$ expansion of
$\mathcal{A}_{tW}^{(1)}$ converges rapidly, too.
The goodness of our best approximation for $\mathcal{A}_{tW}^{(1)}$ may be
estimated by considering its relative deviation from the second best one.
For $M_H=120$~GeV, 140~GeV, and $2 M_W$, this amounts to 0.3\%, 1.0\%, and
2.8\%, respectively.
The situation is very similar to the one in Fig.~\ref{loopW}(a).
In fact, the corresponding figures for $\mathcal{A}_{W}^{(0)}$ are 0.4\%,
1.1\%, and 3.1\%.
We thus expect that the goodness of the approximation of
$\mathcal{A}_{tW}^{(1)}$ by the expansion through $\mathcal{O}(\tau_W^4)$ is
comparable to the case of $\mathcal{A}_{W}^{(0)}$.

\begin{figure*}[ht]
\centering
\begin{tabular}{cc}
\includegraphics[width=90mm]{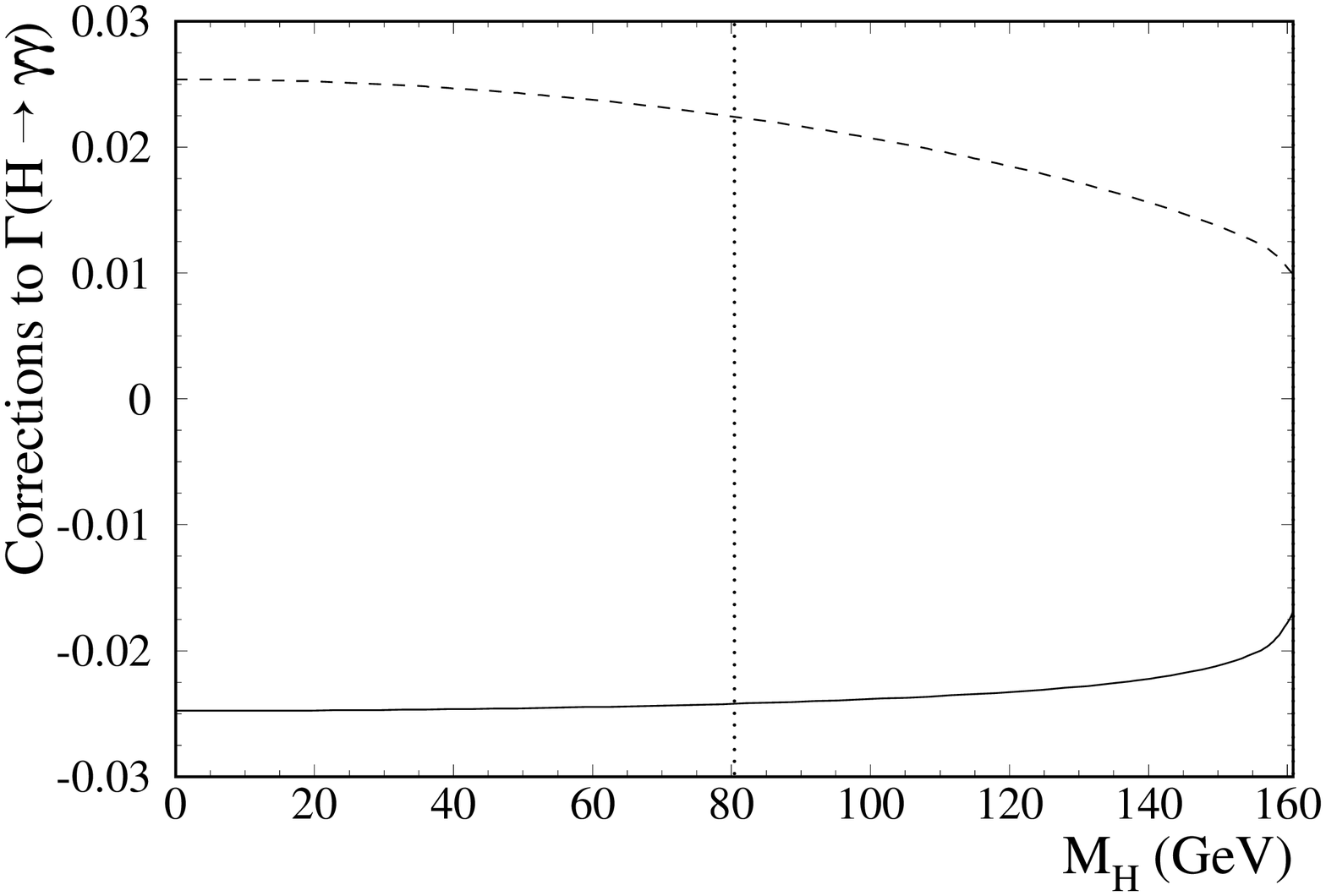} &
\includegraphics[width=90mm]{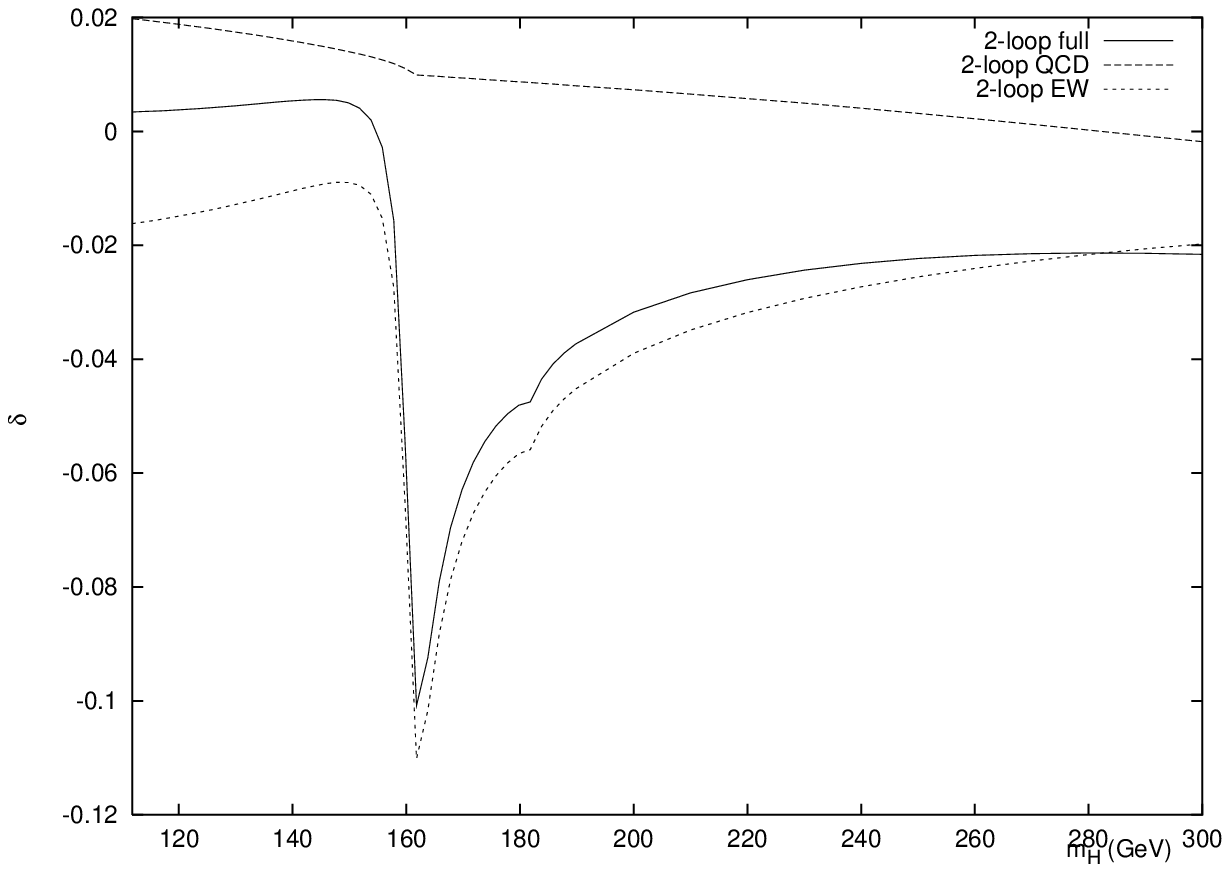} \\
(a) & (b)
\end{tabular}
\caption{(a) $\mathcal{O}(G_F M_t^2)$ \cite{fugel} (solid line),
$\mathcal{O}(\alpha_s)$ \cite{Djouadi:1990aj} (dashed line), and (b)
$\mathcal{O}(n_fG_FM_W^2)$ \cite{Aglietti:2004nj} (dotted line) two-loop
corrections to $\Gamma(H\to\gamma\gamma)$ as functions of $M_H$.}
\label{fig:NLO}
\end{figure*}

For the comparison with future measurements of $\Gamma(H\to\gamma\gamma)$, all
known corrections have to be included in Eq.~(\ref{Notation}).
In this connection, it is interesting to compare the
$\mathcal{O}(G_F M_t^2)$ electroweak correction discussed above with the
well-known $\mathcal{O}(\alpha_s)$ QCD correction \cite{Djouadi:1990aj} and
the $\mathcal{O}(N_fG_FM_W^2)$ electroweak correction induced by light-fermion
loops, which has become available recently \cite{Aglietti:2004nj}.
This is done in Figs.~\ref{fig:NLO}(a) and (b), where the respective
corrections to $\Gamma(H\to\gamma\gamma)$ are displayed as functions of $M_H$.
As in Fig.~\ref{loopW}, the dotted vertical line and the right edge of the
frame in Fig.~\ref{fig:NLO}(a) margin the $M_H$ range $M_W\le M_H\le 2M_W$.
We observe that, within the latter, the $\mathcal{O}(G_F M_t^2)$ correction
slightly exceeds the $\mathcal{O}(\alpha_s)$ one in magnitude, a rather
surprising finding.
Due to the sign difference, the two corrections practically compensate each 
other.
The $\mathcal{O}(N_fG_FM_W^2)$ correction is also negative, but has a slightly
smaller size than the $\mathcal{O}(G_F M_t^2)$ one.

\section{\label{sec:summary}Conclusions}

We discussed the dominant two-loop electroweak correction, of
$\mathcal{O}(G_{F} M_{t}^{2})$, to the partial width of the decay into two
photons of the SM Higgs boson in the intermediate mass range,
$M_W\le M_H\le 2M_W$, where this process is of great phenomenological
relevance for searches at hadron colliders and precision tests at the ILC.

The relevant Feynman diagrams were evaluated with the aid of the
asymptotic-expansion technique exploiting the mass hierarchy
$M_H\ll 2M_W\ll 2M_t$.
In this way, an expansion of the full $\mathcal{O}(G_FM_t^2)$ result in the
mass ratio $\tau_W=M_H^2/(2M_W)^{2}$ through $\mathcal{O}(\tau_W^4)$ was
obtained.
The convergence property of this expansion and the experience with the
analogue expansion at the Born level, where the exact result is available for 
reference, lead one to believe that these five terms should provide a very good
approximation to the exact result for $M_H\alt140$~GeV.
By the same token, the deviation of this approximation for the
$\mathcal{O}(G_FM_t^2)$ amplitude $\mathcal{A}_{tW}^{(1)}$ from the unknown
exact result for this quantity is likely to range from 2\% to 20\% as the
value of $M_H$ runs from 140~GeV to $2M_W$.

In the intermediate Higgs-boson mass range, the $\mathcal{O}(G_FM_t^2)$
electroweak correction reduces the size of $\Gamma(H\to\gamma\gamma)$ by
approximately 2.5\% and thus fully cancels the positive shift due to the
well-known $\mathcal{O}(\alpha_s)$ QCD correction \cite{Djouadi:1990aj}.

As a by-product of this analysis, also the $\mathcal{O}(G_FM_t^2)$ correction
to the partial width of the decay into two gluon jets of the intermediate-mass
Higgs boson was recovered \cite{Djouadi:1994ge}.

\section*{Note added}

After the workshop, a preprint \cite{Degrassi:2005mc} appeared in which the
two-loop electroweak corrections to $\Gamma(H\to\gamma\gamma)$ involving
intermediate bosons and the top quark are computed as expansions in
$q^2/(2M_W)^2$, where $q$ is the four-momentum of the decaying Higgs boson.
In that paper, also the key result of Ref.~\cite{fugel}, Eq.~(\ref{eq:end}),
is confirmed.

\begin{acknowledgments}
The author thanks Frank Fugel and Matthias Steinhauser for their collaboration
on this work.
This work was supported in part by BMBF Grant No.\ 05~HT4GUA/4 and HGF Grant
No.\ VH-NG-008.
\end{acknowledgments}

\end{document}